\documentclass[pre,onecolumn,superscriptaddress,nopacs, nofootinbib,showkeys]{revtex4}

\pdfoutput=1

\usepackage{graphics,amssymb}

\newcommand{\fig}[5]
{
\begin{figure}[#4]
\begin{center}
\resizebox{#5 \columnwidth}{!}{\includegraphics{#1}}
\caption{\label{#3}#2}
\end{center}    
\end{figure}
}

\newcommand{\figstar}[5]
{
\begin{figure*}[#4]
\begin{center}
\resizebox{#5 \columnwidth}{!}{\includegraphics{#1}}
\caption{\label{#3}#2}          
\end{center}        
\end{figure*}
}

\begin{document}

\title{Dynamics on the Way to Forming Glass: Bubbles in Space-time}

\author{David Chandler}
\affiliation{Department of Chemistry, University of California, Berkeley, California 94720}

\author{Juan P. Garrahan}
\affiliation{School of Physics and Astronomy, University of Nottingham, Nottingham NG7 2RD, U.K.}

\date{\today}

\begin{abstract}
We review a theoretical perspective of the dynamics of glass forming liquids and the glass transition.  It is a perspective we have developed with our collaborators during this decade.  It is based upon the structure of trajectory space.  This structure emerges from spatial correlations of dynamics that appear in disordered systems as they approach non-ergodic or jammed states.  It is characterized in terms of dynamical heterogeneity, facilitation and excitation lines. These features are associated with a newly discovered class of non-equilibrium phase transitions.  Equilibrium properties have little if anything to do with it.  The broken symmetries of these transitions are obscure or absent in spatial structures, but they are vivid in space-time (i.e., trajectory space).  In our view, the glass transition is an example of this class of transitions.  The basic ideas and principles we review were originally developed through the analysis of idealized and abstract models.  Nevertheless, the central ideas are easily illustrated with reference to molecular dynamics of more realistic atomistic models, and we use that illustrative approach here.     
\end{abstract}


\maketitle

\tableofcontents

\section{Introduction}

This article reviews current understanding of highly correlated dynamics characterizing structural glass forming liquids -- super-cooled liquids approaching the glassy state.  It is a hotly contested topic \cite{Quote-PWA,Quote-Science,Quote-NYT}, and many recent reviews have been written about it, for example \cite{Review-Angell,Review-Ediger-et-al,Review-Angell-et-al,Review-Debenedetti,Review-Ediger,Review-Wolynes,Review-Cavagna}.  The majority of this writing adopts one of two perspectives:  a thermodynamic view, the origins of which are often attributed to Goldstein \cite{Goldstein} and to Adam \& Gibbs \cite{Adam-Gibbs}, and a small-fluctuation dynamical view called ``mode coupling theory'' (MCT) pioneered by G\"otze \cite{MCT1,MCT2,MCT3}.  Syntheses of these two approaches have been proposed \cite{KTW,Xia-Wolynes,Schweizer-Saltzman}, and some plausible arguments link aspects of such syntheses to dynamical issues we discuss in this article \cite{BB}.  By and large, however, what we describe has a different basis, originating with discoveries that structural glass forming liquids exhibit pronounced dynamic heterogeneity.  

These discoveries were made with both experiment \cite{Spiess,Cicerone-Ediger,Sillescu,Weeks-et-al,Kegel-vanBlaaderen,Richert} and computer simulation \cite{Miyagawa-et-al,Kob-et-al,Yamamoto-Onuki,Perera-Harrowell,Glotzer-review,HCA-review}.  Dynamic heterogeneity emerges spontaneously from dynamics.  It is a fluctuation dominated phenomenon that is largely independent of thermodynamics.  Thermodynamics pertains mainly to mean behaviors.  Mode coupling theory accounts for some fluctuation effects, but it is reliable only up to an onset of fluctuation dominance.  As we describe in these pages, properties that distinguish structural glass formers from other materials are captured by physical models with large dynamic heterogeneity, irrespective of average thermal properties, and irrespective of many molecular details.

Our presentation is qualitative, at the level of that found in the textbooks by Chandler \cite{Chandler} and by Barrat \& Hansen \cite{BH}. We leave technical discussions of theoretical techniques and analyses to a future review \cite{APreview}.  Much of what we present here draws on results from numerical simulation of a simple atomistic model of a glass forming material.  Our perspective on what occurs in that model and how it relates to experiment originated in theoretical work on idealized lattice models. These models are simple enough that they can be analyzed in detail, and remarkably, despite their simplicity, they successfully imitate many aspects of real super-cooled liquids, and many pertinent principles can be derived.

Model building and solving is a traditional and important approach in statistical mechanics. 
Inarguable results for non-trivial models provides the basis for dismissing poor approximations, and for building the foundations of correct and general statements.  For the topic of structural glass formers, the origins of the most used and useful models can be traced to Andersen and his co-workers \cite{FA,Kob-Andersen-KCM,KALJ}.  One class is a collection of the so-called ``kinetically constrained models" (KCMs) \cite{FA,Kob-Andersen-KCM,Jackle-Eisinger,Jackle-Kronig,KCMs}, which are lattice models with Markovian dynamical rules that codify principles of facilitated \cite{Glarum,Philips} and hierarchical \cite{Palmer} dynamics.  The other class is composed of continuum models of fluid mixtures, which are atomistic models that can be simulated on computers \cite{KALJ,HCA-review}.  At a superficial level, the two classes appear to be very different, but for their glassy behaviors they are similar. 

Approaching the end of this review, we reach the point where we are able to describe the recent discovery that models of structural glass formers exhibit a first-order phase transition between ergodic melt and non-ergodic glass phases \cite{Order-disorder}.  It is a non-equilibrium phase transition, one controlled by fields that couple to trajectories -- paths through state space.  It is not controlled by thermodynamic fields like temperature and pressure, which couple to only states.  Indeed, while only vague precursors are found in equilibrium measurements, a sharp unambiguous transition is found in systems driven away from equilibrium.  This finding allows us to develop principles of the glass transition in terms of familiar concepts like order parameters, broken symmetries and wetting.

\section{Disordered Condensed Matter}

Glasses are solids without evident structural order.  They are formed from super-cooled liquids that slowly relax to equilibrium, sufficiently slowly that an arrested disordered material can be formed before an ordered crystal can assemble \cite{Review-Angell,Review-Ediger-et-al,Review-Angell-et-al,Review-Debenedetti,Review-Ediger,Review-Wolynes,Review-Cavagna}.  Figure~\ref{fig:tExp} illustrates this juxtaposition of time scales.  The transition to a glass is a dynamic process that occurs continuously over a range of temperatures, and the range depends upon the method by which the melt is cooled.  The materials thus formed are truly solids, as much so as their ordered counterparts.

Glasses do age and change their properties, some faster than others. Sometimes, this aging is the result of structural transformations, not fluid flow. For instance, initially clear kitchen glass wear may become foggy, and initially elastic organic glass may become brittle.  These are examples of physical aging.  The first is due to domains of quartz crystal growing within otherwise disordered regions.  This phenomenon is accelerated through repeated heating cycles in a dishwasher.

Solids, unlike liquids, are non-ergodic phases -- atoms are fixed at or close to their initial positions for macroscopically long periods of time.  In the case of crystals, the reason for this behavior seems obvious in view of the material's atomic structure.  The energy and entropy are such that an ordered dense array of atoms is favorable, and an atom cannot move to a neighboring position without a massive rearrangement of atoms.  The free energy for such a rearrangement would be formidable.  Any appreciable atomic motions, therefore, must be due to defects that can move with lower free energy cost.  Since a glass appears to be filled with structural defects, why is it also a solid?

The answer that most often applies is appreciated by noting how the compression of a disordered system of hard space-filling objects may eventually reach a point of random close packing \cite{Bernal, Torquato}.  This compression is sometimes referred to as ``jamming'' \cite{Jam}.  The resulting system is non-ergodic and therefore a glass.  In a jammed or nearly jammed system, rare configurations that permit particles to move can represent only a small subset of deviations from crystal order. These rare configurations are the local dynamical excitations of a glass or glass former, excitations that lead to aging or relaxation.  

Behaviors of systems approaching the point of jamming can be observed in granular materials, and these behaviors have commonalities with those of structural glass formers \cite{Goldman,Dauchot,Keys}.  In particular, restrictive forces between molecules in a super-cooled liquid lead to glassy behavior, and at high enough packing fractions the most important of these forces are repulsive intermolecular interactions.  This idea, that the molecular structure and dynamics of dense disordered systems is often dominated by repulsive forces forms the basis for understanding the molecular behaviors of most normal liquids \cite{Jonas,WCA1}, its origins can be traced to van der Waals \cite{VanderWaals}, to Bernal \cite{Bernal} and to Widom~\cite{Widom}, and it underlies the standard equilibrium theory of simple liquids \cite{WCA2,Hansen-McDonald}.

\fig{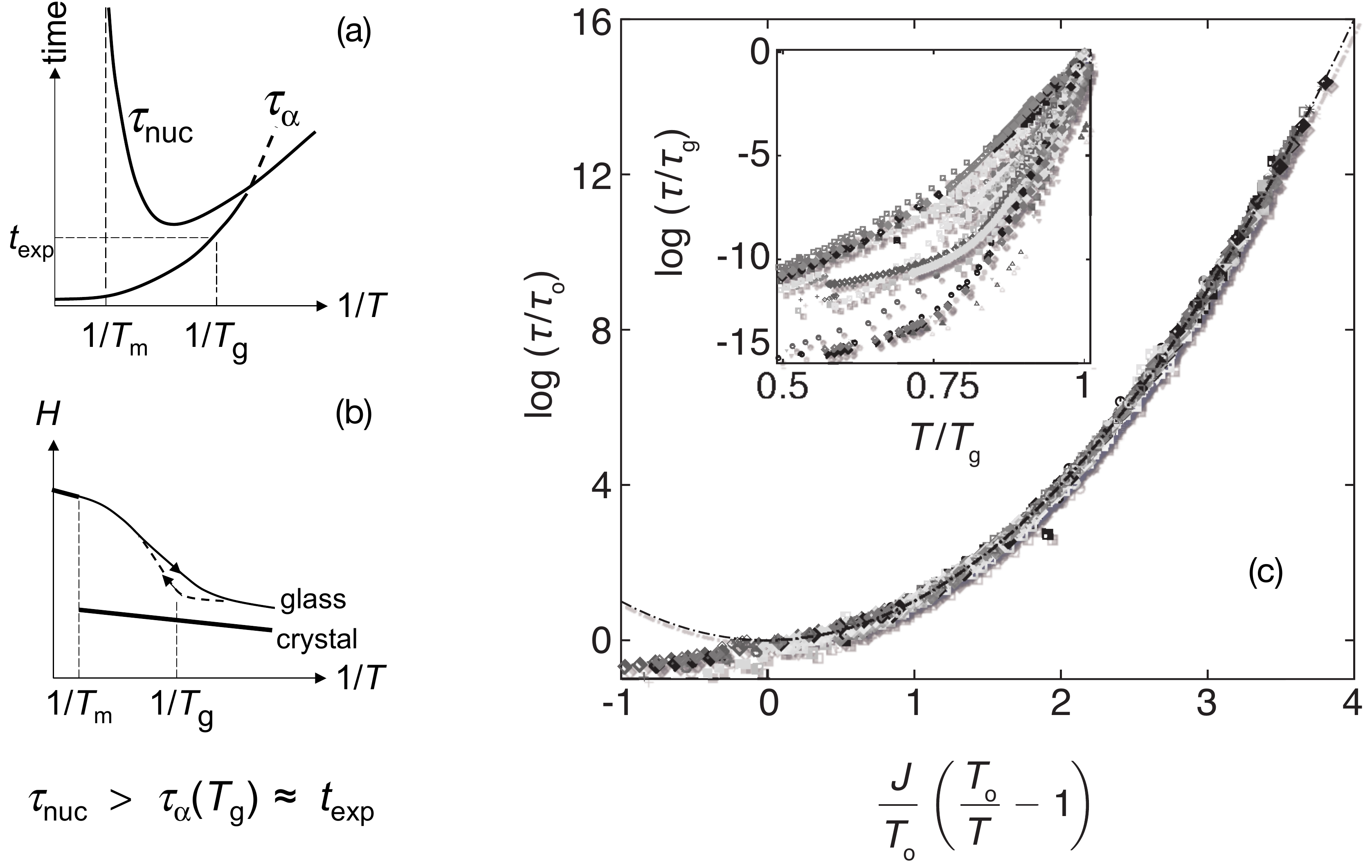}{Temperatures and time scales for making glass.  Panel (a) illustrates the time to nucleate and grow a crystal from its melt, $\tau_{\mathrm{nuc} }$, and the time to relax the structure of a melt, $\tau_{\alpha}$.  The former is a non-monotonic function of temperature $T$~\cite{OnukiBook}.  It is infinite at the melting temperature, $T_{\mathrm{m}}$, and below that point it initially decreases before entering the regime where crystal growth is dominated by ripening; at that stage, its kinetics is limited by diffusion, and $\tau_{\mathrm{nuc}}$ then increases with decreasing $T$~\cite{Ediger-nucleation}. In contrast, $\tau_{\alpha}$ increases monotonically with decreasing $T$.  The two times may intersect.  The material will become a glass if the time scale over which the material can be observed and manipulated experimentally, $t_{\mathrm{exp}}$, is shorter than $\tau_{\mathrm{nuc}}$.  The glass transition temperature, $T_{\mathrm{g}}$, is a temperature at which the structural relaxation time is about the same as the limiting experimental time.  $T_{\mathrm{g}}$ is often taken to be the temperature at which the relaxation time $\tau_{\alpha}$ becomes $10^2$s \cite{Review-Angell,Review-Ediger-et-al,Review-Angell-et-al}.   $T_{\mathrm{g}}$ is usually below the equilibrium melting temperature, $T_{\mathrm{m}}$, and $T_{\mathrm{g}}$ depends upon both the system and the experimental time limits and protocols; for most observed glass formers $T_{\mathrm{g}}$ is about $2/3$ of $T_{\mathrm{m}}$ \cite{Review-Angell,Review-Ediger-et-al,Review-Angell-et-al}.  Panel (b) illustrates typical temperature dependence of the enthalpy, $H$, for a liquid, its crystal and its glass.  Curves of this type are constructed experimentally by recording the system's enthalpy as it is cooled or warmed at a rate of about $1/t_{\mathrm{exp}}$.  If a glass is formed, hysteresis occurs near $T=T_{\mathrm{g}}$.  The temperature at which the warming curve will have an inflection point is sometimes defined as the glass transition temperature.  (c) Collapse of equilibrium transport data for fragile liquids in terms of the relaxation time at the onset to fluctuation dominance, $\tau_{\mathrm{o}}$, the temperature at that onset, $T_{\mathrm{o}}$, and a single energy parameter, $J$. 
The quantity $\tau$ stands for either a structural relaxation time, $\tau_{\alpha}$, most often determined from dielectric measurements, or the shear viscosity, $\eta$, determined from transport measurements.  The quantity $\tau_{\mathrm{o}}$ refers to this same property at the onset temperature.   This is an application of Eq.\ (\ref{TauFragile}) to more than 1000 data points from 67 different systems.  Adapted from \cite{Elmatad}. 
Inset shows the same data graphed in a so-called ``Angell'' plot \cite{Review-Angell}.}{fig:tExp}{ht}{.8}

\section{Repulsive and attractive forces in condensed matter}

Attractive branches of typical intermolecular potentials are responsible for making a dense liquid stable at standard pressures, but their effects are those of a mean or background potential that contribute little to energy fluctuations and forces that govern microscopic arrangements and movements of atoms \cite{WCA1, Widom}.  These movements and arrangements are much like those of a vibrating box of marbles or ball bearings.  If the packing fraction is not too high, each atom can wiggle and jiggle and diffuse. Such random thermal motions are not rare (i.e., not activated), and they do not involve correlated movements of many particles.  At a high enough packing fraction, however, the space available to each particle is constricted, and any significant displacement of one atom must be accompanied by rearrangements of several neighboring atoms.  Examples of these correlated motions are illustrated below when we discuss dynamic heterogeneity.  

Exceptions to where forces other than repulsions dominate are called ``associated'' liquids and ``networked'' glasses.  Water is an associated liquid.  Strong hydrogen bonds significantly influence its structure and dynamics.  Most liquids do not possess such large attractive intermolecular forces.   Hydrogen bonding causes liquid water molecules to arrange in a disordered tetrahedral network \cite{Stillinger-water}.  On small length scales, the liquid appears much like its ordered crystal ice.  It is a relatively open structure.  Entropic forces oppose it because it makes inefficient use of configuration space.  On raising temperature, entropy quickly wins the competition with attractive hydrogen bonding.  The preference for tetrahedral arrangements of water molecules ceases above about 50$^{\mathrm{o}}$C, well below the boiling temperature.

Window glass is a networked glass.  It is composed mostly of SiO$_2$ (i.e., silica), in which atoms interact with highly directional potentials that, like water, favor tetrahedral bonding of neighbors.  Whether due to strong directional attractive forces, like those in silica, or due repulsive intermolecular forces that dominate in organic liquids, glassy dynamics and non-ergodicity are the results of molecular confinement -- a constriction of accessible configuration space, and the concomitant sparsity of regions where molecular reorganizations are possible.   
This constriction is termed ``frustration" \cite{Binder-Young}.  In particular, part of the potential energy surface kinetically blocks or frustrates access to ordered lowest energy states, limiting configurations to higher energy disorganized configurations, of which there are many.  In the case of organic glass, it is the irregular shapes of the mutually excluding molecules that create this frustration; in the case of networked glass, it is the directionality of attractive forces that create this frustration.

While typical attractive forces in dense non-associated liquids cannot compete with repulsive packing forces, they can still have large effects.  In particular, when molecular packing limits available configuration space to the point where structural reorganization involves coordinated displacements of several particles, many small attractive forces between separate pairs of molecules will act in the same direction and thus add significantly to the net forces controlling dynamics.  In this way, attractive forces can enhance, not oppose, the effects of correlated particle motions.  This is well known and understood in the context of equilibrium phenomena, where the presence of a large enough length scale heterogeneity, attractions are large contributors to the resulting unbalanced potentials between particles~\cite{WeeksAnnRev, ChandlerNature}.   
In glass forming liquids, the source of heterogeneity is correlated dynamics, which we will soon discuss in some detail.  Nevertheless, in glass formers it is the repulsions that are the key frustrating forces, and the source of heterogeneity is correlated dynamics, which we will soon discuss in some detail.

The terminology ``packing fraction'' refers to the fraction of space occupied by the molecular volumes of the particles in the system.  For example, the packing fraction of an ordered array of close-packed hard spheres is $(\pi / 6) \rho \sigma^3 = \pi /3 \sqrt{2}\approx0.7 $, where $\rho$ is the number of spheres per unit volume, and $\sigma$ is the sphere diameter.  Close packing of random hard spheres is estimated to be at about 10\% lower density than that of ordered hard spheres \cite{Aste}.  An equilibrated fluid of hard spheres freezes into its crystal when compressed to packing fraction of about 0.5, and the crystal's density is about 10\% larger than that of the coexisting fluid \cite{Wood,Alder}.  Ordinary freezing of simple liquids is a reflection of this athermal first-order phase transition of hard spheres \cite{Widom,Laird}.  

In real systems, repulsive forces are softer than hard cores, and as a result, the parameters characterizing molecular space filling volumes are functions of temperature \cite{WCA1,AWC}.  Roughly speaking, two atoms can approach each other only to the point where their repulsive potential of interaction reaches about $k_{\mathrm{B}} T$ ($k_{\mathrm{B}}$ stands for Boltzmann's constant).  Thus, the typical distance of closest approach grows with decreasing temperature, so that at fixed molecular density, $\rho$, the packing fraction increases with decreasing $T$.  At constant pressure with decreasing temperature, there can be further increase in packing fraction because the density $\rho$ will increase when the coefficient of thermal expansion is positive.

Thus, it is clear from the context of intermolecular forces why super-cooling a liquid can produce constricting effects that slow dynamics and eventually trap it in a glassy state.  But what is the nature of that transition? Figure~\ref{fig:tExp} depicts what occurs with respect to changes in thermodynamic variables like temperature and pressure.  At low enough temperatures (or high enough packing fractions), one observes hysteresis in thermal properties like the enthalpy.  This behavior occurs smoothly, and it depends upon the rates at which the system is cooled and warmed (or compressed and expanded).  It reflects intrinsic non-linear dynamics.  For condensed matter, such non-linearity is generally associated with correlations between many degrees of freedom, and strong enough correlations produce order-disorder transitions like the sharp equilibrium transition between liquid and crystal.  But the formation of glass exhibits nothing like this singular behavior, at least with respect to changes in thermodynamic variables like temperature and pressure.  To reinforce this point, it is helpful to consider the experimental temperature dependence of structural relaxation, which we turn to now.

\section{Growing time scales -- universal, not singular and not thermodynamic} 

Structural glass forming liquids are generally catalogued as being either ``strong'' or ``fragile.''  This terminology is due to Angell~\cite{Review-Angell}.  Strong materials are those for which measured equilibrium transport times grow with lowering temperature $T$ in an Arrhenius fashion, 
\begin{equation}
\label{TauStrong}
\log \left(\tau/\tau_0\right)_{\mathrm{strong}} = (E^*/k_{\mathrm{B}})(1/T - 1/T_0) .
\end{equation}
Here, we use $\tau$ to stand for a structural relaxation time, $\tau_{\alpha}$, or a surrogate like viscosity, and $T_0$ denotes a temperature at some reference point.  When transport properties are well fitted by this form, provided $E^*/k_{\mathrm{B}}T\gg1$, one may conclude that there is single type of event that leads to the longest time relaxation, and that the energy required to activate that event is $E^*$~\cite{Chandler}.

Fragile materials are those that are super-Arrhenius -- their measurable relaxation times grow faster than exponential in $1/T$.   Most glass formers are fragile over the range of temperatures where they can be equilibrated (and thus observed to relax).  The experimental data for these materials collapse to a form only slightly more complicated than Arrhenius, namely quadratic in reciprocal temperature~\cite{Elmatad},
\begin{equation}
\label{TauFragile}
\log \left(\tau/\tau_{\mathrm{o}}\right)_{\mathrm{fragile}} = \left(J/T_{\mathrm{o}} \right)^2 \left(T_{\mathrm{o}}/T - 1 \right)^2,
\end{equation}  
where $k_{\mathrm{B}}J$ is an energy determining the rate of growth of $\tau$ as temperature is decreased from a reference temperature, $T_{\mathrm{o}}$, which is called the ``onset temperature.'' This temperature marks the onset to rapid growth of $\tau$ upon lowering $T$.  Transport properties of normal liquids have relatively insignificant temperature dependence, and this weak dependence mostly reflects the coefficient of thermal expansion, not activated dynamics~\cite{Jonas,WCA1}.  As such, the onset temperature is easily identified from data.   Figure~\ref{fig:tExp}(c) illustrates the collapse of data below the onset temperature \cite{Elmatad}.

Temperature dependences of transport properties are frequently fit to the Vogel-Fulcher-Tamman (VFT) formula, $\log \left( \tau/ \tau_{0} \right)_{\mathrm{VFT}}=  A/(T-T_{\mathrm{VFT}})$ \cite{Review-Angell,Review-Ediger-et-al,Review-Angell-et-al}.  Here $A$, $\tau_0$ and $T_{\mathrm{VFT}}$ are the three constants used to fit data to this form.  These fitting constants are not directly related to properties of measurable reference points.  For example, the temperature $T_{\mathrm{VFT}}$ is the temperature at which $\tau$ is imagined to diverge, hence the system cannot be observed to relax at that temperature.  Despite this metaphysical quality, and despite the fact that experimental data make no compelling case for it \cite{Elmatad, Dyre}, the VFT expression is sometimes referred to as a ``law".

\subsection{Dynamical perspective}

Super-Arrhenius behavior results from a distribution of relaxation times contributing to the longest time relaxation.  For example, if the relaxation time $\tau_{\mathrm{o}}\exp[\Delta (1/T - 1/T_{\mathrm{o}})]$ is random because the activation energy $\Delta$ is random, averaging over $\Delta$ can yield the quadratic function used to collapse data in Fig~\ref{fig:tExp}.  In that case, $\Delta$ would obey Gaussian statistics with a mean of zero and a variance equal to $2J^2$~\cite{Bassler}.   The super-Arrhenius behavior of Eq.\ (\ref{TauFragile}) also emerges from lattice models where microscopic reversible dynamics is hierarchical, with time scales for relaxing domains growing logarithmically with domain size \cite{SollichEvans}.  Later, we will see hierarchical behavior in the dynamic heterogeneity of an atomistic model. 

As temperature is lowered, typical separations between mobile relaxing regions become larger, so that with hierarchical dynamics, the number of coordinated steps required to relax an inactive region will also be larger.  As a result, the apparent activation energy will grow with decreasing $T$.  This growth will ultimately produce an activation energy that is larger than that required to simply avoid dynamical constraints, and in the absence of constraints dynamics cannot be hierarchical.  Thus, when mobile regions are sufficiently sparse, i.e., when temperature is sufficiently low, we anticipate a crossover from super-Arrhenius to Arrhenius behavior, where the single pertinent activation barrier in the latter case is the energy required to disrupt the constraining network of forces.  Lattice models predict this fragile-to-strong behavior \cite{xover1,xover2,pnas}.  Experimental observations of such a crossover have been reported for confined super-cooled water \cite{Chen}, and for polymeric melts \cite{26f}, but the investigators attribute the crossover in the former case to a hypothesized equilibrium liquid-liquid phase transition, and in the latter case to supposed non-equilibrium phenomena.  Clarification will require further experimental work.

\subsection{Thermodynamic perspective}

The VFT formula is often discussed in the context of the Adam-Gibbs expression, $\log \left( \tau/ \tau_{0} \right)_{\mathrm{AG}} \propto 1/TS_{\mathrm{con}}$, where $S_{\mathrm{con}}$ is called ``configurational entropy''~\cite{Adam-Gibbs}.  $S_{\mathrm{con}}$ is not uniquely defined, but it is usually equated with the molar entropy of the super-cooled liquid less that of the crystal at the same temperature.  This entropy difference can vanish, and the temperature at which it vanishes is Kauzmann's temperature, $T_{\mathrm{K}}$~\cite{Kauzmann, StillingerTk}.  This would suggest that $S_{\mathrm{con}}$ vanishes at $T=T_{\mathrm{K}}$ and that $\tau$ would therefore diverge as $T \rightarrow T_{\mathrm{K}}$.  Kauzmann's temperature would then be the temperature of an ideal glass transition -- a non-zero temperature that would be unapproachable at equilibrium conditions.  There is a convincing theoretical argument contradicting this metaphysical interpretation of $T_{\mathrm{K}}$~\cite{StillingerNoTk}.  Yet, by adopting this approximation and expanding to linear order in $T-T_{\mathrm{K}}$, we have $S_{\mathrm{con}} \approx  [(T-T_{\mathrm{K}})/T_{\mathrm{K}}]\,\Delta C$, where $\Delta C$ is the difference between liquid and crystal molar heat capacities.  The Adam-Gibbs expression then becomes the VFT formula, with $A \propto T_{\mathrm{K}}/\Delta C$, and $T_{\mathrm{K}}=T_{\mathrm{VFT}}$, thus suggesting a link between equilibrium thermodynamics, diverging time scales and the glass transition.  Experimental support for this link is mixed \cite{Angell-Sc, Tanaka, Example,McKenna}, at least in part because applications require extrapolations of experimental data that exhibit hysteresis. 

Nevertheless, proposals relating the VFT formula and other divergent expressions to extrapolated thermodynamic properties are deeply entrenched in the literature on glassy physics.  On theoretical grounds, this entrenchment may reflect results derived from mean field models of spin glasses \cite{KTW,SpinGlass}, which display both dynamical and thermodynamical singularities.  But static disorder is a given rather than an emergent property of spin glasses, so a transition that might appear in spin glasses need not appear in structural glass formers.  
Not discounting the important practical role of contrasting and correlating thermodynamic and dynamic properties~\cite{Richert98,Wang}, there are too many unexplained exceptions to be able to use these correlations as the basis for a rigorous analysis.

\section{Order parameters and two (or more) step relaxation}

While the previous section emphasizes that there is nothing singular to be found in equilibrium
data, there is in fact an order-disorder phenomena connected to the glass transition, but it occurs away from equilibrium, and it is not entirely controlled by thermodynamic
variables like temperature and pressure. Explaining why requires some discussion, the first
step of which is to consider appropriate order parameters.

Just as the volume per molecule distinguishes two fluid phases -- liquid and gas -- there should be a characteristic property that distinguishes liquid and glass.  A distinction based upon molecular configurations, however, is not obvious because liquids and glasses have similar if not identical densities and local structures.  The most obvious difference between the two is dynamical.  Normal and supercooled liquids are ergodic while glass is non-ergodic.  An acceptable order parameter should highlight this distinction.  For example, one measure of local dynamical activity is the squared displacement of a particle's position between time $t$ and time $t+\Delta t$, the quantity $|\mathbf{r}_i(t+\Delta t)-\mathbf{r}_i(t)|^2$, where $\mathbf{r}_i(t)$ is the position of the $i$th particle at time $t$.  At equilibrium, its mean value is independent of particle label and time origin, $\left< |\mathbf{r}_i(t+\Delta t)-\mathbf{r}_i(t)|^2\right> =\left< |\mathbf{r}_j(\Delta t)-\mathbf{r}_j(0)|^2 \right>$, where $i \neq j$.  For large enough $\Delta t$,  $\left< |\mathbf{r}_j(\Delta t)-\mathbf{r}_j(0)|^2 \right> \to 2dD \Delta t$, where $D$ is a particle's self diffusion constant and $d$ is dimensionality.  The pointed brackets, $\left<...\right>$, indicates equilibrium averaging over conditions at some point in time, or equivalently (assuming ergodicity) averaging over $t$ for a very long trajectory. 

While a single-particle transport property like the self diffusion constant is a measure of dynamical activity, it is not by itself a suitable order parameter, which must be extensive in both space and time -- extensive in space (or number of molecules) because we wish to distinguish systems viewed over macroscopic length scales (and thus detect broken spatial symmetry), and extensive in time because we wish to distinguish systems viewed over long observation times (and thus detect broken time symmetry).  Such an order parameter can be formed by summing the local measure over the entire system for the entire time of observation (assuming both are very large).  For example, one possible order parameter could be
\begin{equation}
\label{Activity_K}
K = \Delta t \sum_{t=0}^{t_{\mathrm{obs}}} { \sum_{i=1}^{N} {|\mathbf{r}_i(t+\Delta t)-\mathbf{r}_i(t)|^2}}\,,
\end{equation}
where $N$ stands for the total number of particles considered, and $t_{\mathrm{obs}}$ stands for the total amount of time the system is observed, incremented in the sum by steps of $\Delta t$.  
Another possible choice is
\begin{equation}
\label{Overlap_Q}
Q = \Delta t \sum_{t=0}^{t_{\mathrm{obs}}} { \sum_{i=1}^{N} {\exp \{i \mathbf{k}\cdot \left[\mathbf{r}_i(t+\Delta t)-\mathbf{r}_i(t) \right] \}}}\,,
\end{equation}
which is a measure of overlap of configuration space at time $t$ with that at time $t' = t+\Delta t$, resolved on length scale $2\pi / k$.  When $2\pi/k$ is a molecular length,  extensive overlap for very large $\Delta t$ is a signature of a non-ergodic solid \cite{Kob-Binder}.

For the case of equilibrium dynamics, the mean values of these two order parameters are given by standard equilibrium correlation functions.  In particular,
\begin{equation}
\label{MeanActivity}
\left<K\right>= N t_{\mathrm{obs}} \left<|\mathbf{r}_1(t+\Delta t)-\mathbf{r}_1(t)|^2 \right> = 2dN t_{\mathrm{obs}}D \Delta t ,
\end{equation}
where the second equality is true for $\Delta t$ large enough for translational motion to be diffusive, and
\begin{equation}
\label{VanHove_F}
\left< Q \right> = Nt_{\mathrm{obs}}F\left( k, \Delta t \right),
\end{equation}
where $F(k, \Delta t) = \left< \exp \{i \mathbf{k}\cdot \left[\mathbf{r}_i(\Delta t)-\mathbf{r}_i(0) \right] \} \right>$ is Van Hove's so-called ``self correlation function'' or ``intermediate scattering function'' \cite{Hansen-McDonald}.

\fig{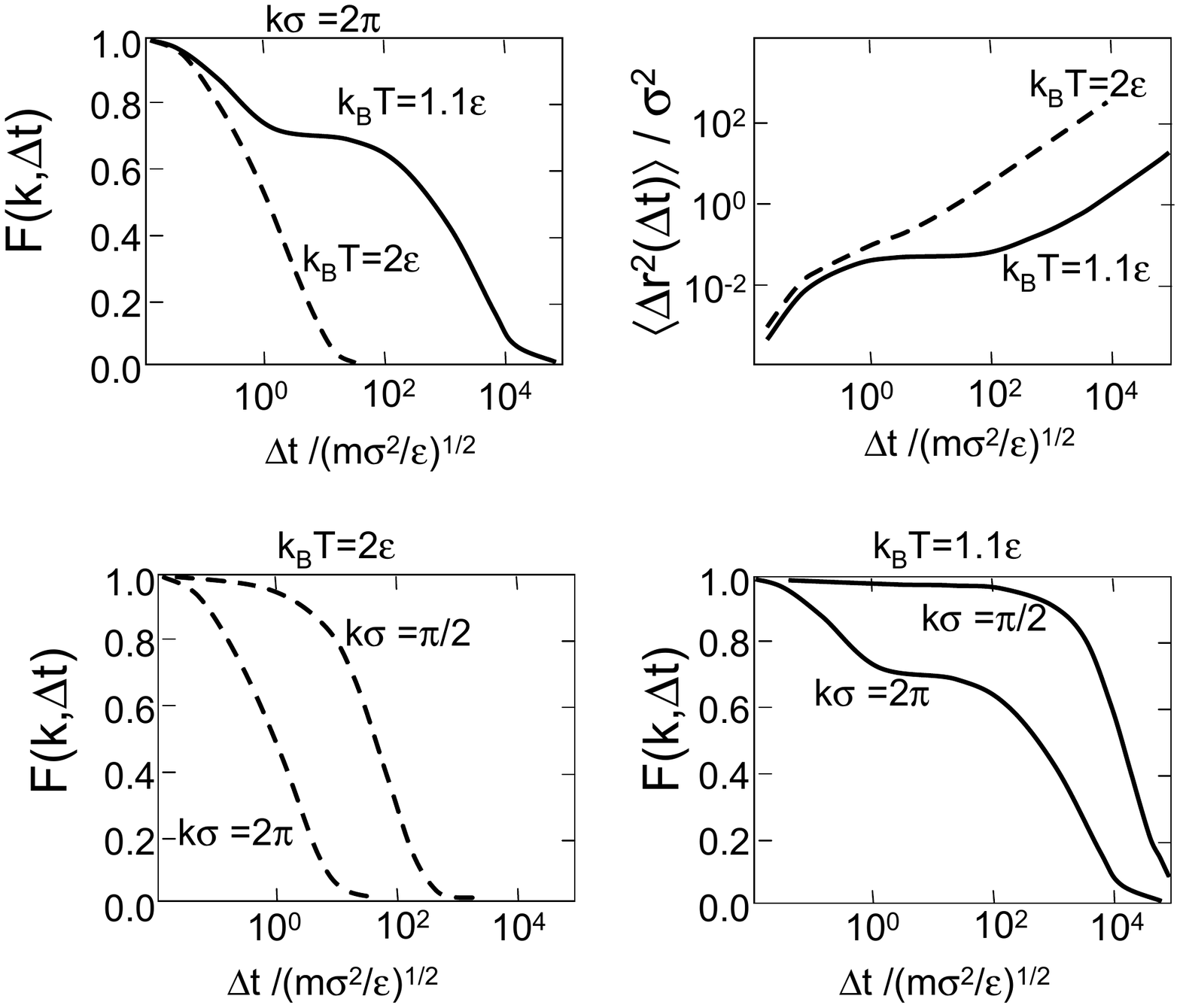}{Equilibrium single-particle time-correlation functions for a two-dimensional glass forming mixture.  The system is composed of 50\% A-particles and 50\% B-particles, which interact with WCA pair potentials \cite{WCA2}.  The length parameter for the A-particles is the unit of length, $\sigma$, while that for the B-particles is $1.4\sigma$.  The pictured correlation functions refer to the larger of the two types of particles.  (Those for the smaller particle behave similarly.)  The net packing fraction of the mixture is $ \pi \rho \left(\sigma_{A}^2 + \sigma_{B}^2 \right)/8 = 0.872$.  All particles have the same mass, $m$, and the same energy parameter, $\epsilon$.  $F(k, \Delta t)$ is the Van Hove correlation function defined in Eq.\ (\ref{VanHove_F}), and $\left<\Delta r^2(\Delta t) \right>$ is an abbreviation for a tagged particle's mean-square displacement $\left< |\mathbf{r}_1(\Delta t)-\mathbf{r}_1(0)|^2\right> $.  Each of the upper two panels juxtapose behaviors above and below the onset to glassy dynamics.  The bottom two panels juxtapose a measure of differing length scale dependences for dynamics above the onset temperature (bottom left) and below the onset temperature (bottom right). }{fig:Corrfuncs}{ht}{.8}

The terminology ``equilibrium dynamics'' refers to trajectories that are time reversal symmetric, preserve an equilibrium distribution of states, and, in the absence of external driving forces, reach this distribution from any initial condition.  In this sense, a super-cooled liquid can be an equilibrium system.  A glass, on the other hand, is the result of non-equilbrium dynamics, and we will see that its creation is the result of broken time-translation symmetry.

Behaviors of the equilibrium correlations functions, and thus the equilibrium mean values of the order parameters, are illustrated in Fig.~\ref{fig:Corrfuncs}.  The specific correlation functions shown are those computed by molecular dynamics simulation for a model $d=2$ fluid mixture composed of classical particles interacting with the repulsive Weeks-Chandler-Andersen (WCA) potentials, i.e., the repulsive branches of Lennard-Jones potentials \cite{WCA1, WCAmix}.  Crystallization of this system is frustrated by the differing sizes of the two components in this mixture, one being 40\% larger than the other.  As such, it is possible to observe equilibrium dynamics in super-cooled fluid states of this system.  The behaviors shown in Fig.\ \ref{fig:Corrfuncs} are typical of results obtained for any number of other such models, in dimensions two and three, that have been studied numerically~\cite{Kob-Binder}.

Figure~\ref{fig:Corrfuncs} shows that a super-cooled glass former relaxes to equilibrium in stages.  The very first stage is a fast inertial relaxation found in all fluids, even dilute gases.  The next stage is the approach to a plateau that persists for more than an order of magnitude, at least as we see it at the lower temperature in the mean-square displacement and in $F(k,\Delta t)$ for $k\sigma = 2\pi$.  Notice, however, that this second stage is not evident at the lower temperature when coarse graining over length scales four time larger than $\sigma$, and it is not evident at the higher temperature.  The wave-vector dependence of $F(k,\Delta t)$ therefore manifests some sort of heterogeneity in dynamics that is present at the lower temperature but not present at the higher temperature.  The last stage of relaxation is the final decay of $F(k,\Delta t)$.  This final decay provides an estimate of a structural relaxation time, namely the $1/e$-time for the Van Hove Function.  We denote this time by $\tau_e(k)$, i.e., $1/e = F\left(k,\tau_e(k)\right)$; and we use simply $\tau_e$ when $k=2\pi / \sigma$, this value of $k$ being roughly the location of the principal peak of the structure factor -- the spatial Fourier transform of the equilibrium pair correlation function.  

While not graphed here in a way to highlight the functional from by which $F(k,\Delta t)$ decays with time $\Delta t$, it is worth noting that at the lower temperature and $k=2\pi/\sigma$, the final decay of the Van Hove function is a stretched exponential.  But the stretching exponent depends upon length scale, so that for very small $k$, $F(k,\Delta t)$ decays with the normal exponential form characteristic of diffusive motion, $\exp (-k^2D\,\Delta t)$.  This multifaceted behavior is a manifestation of fluctuation effects that we describe in the next section.

The last stage of relaxation is called ``$\alpha$-relaxation.''  For a given glass former, structural relaxation occurs in a variety of ways, and there is a broad distribution of structural relaxation times.  The wave-vector dependence of $\tau_e(k)$ is one indication of this fact.  The distribution of structural relaxation times can be usefully partitioned into at least two distinct classes of processes (see Fig.\ \ref{fig:ExVsPer} and discussion surrounding it below).  Irrespective of that analysis, there is a mean value for this distribution, and this mean is often termed \emph{the} alpha relaxation time, $\tau_{\alpha}$.  $\tau_e$ can be a reasonable estimate of $\tau_{\alpha}$, but it is not the only possibility.  Depending upon the physical issues at hand, $\tau_e$ may or may not be the best estimate. 

The prior stage of relaxation, the second stage described above, is called ``$\beta$-relaxation'' (or ``fast" $\beta$-relaxation to distinguish it from the slower ``Johari-Goldstein" $\beta$-processes observed in dielectric relaxation spectra \cite{Review-Angell-et-al}).   The inertial and $\beta$-relaxations occur in both ergodic and non-ergodic phases---in crystals and glasses, as well as in equilibrated super-cooled structural glass formers.  These relaxations reflect the residual dynamics of a rigid material.  Unlike $\beta$-relaxation, $\alpha$-relaxation is absent from non-ergodic phases.  In other words, non-ergodicity -- broken time symmetry and rigidity for all times, what we describe as an ``inactive'' phase -- is manifested by a finite value of $\left<Q\right>/Nt_{\mathrm{obs}}=F(k,\Delta t)$ for $\Delta t \rightarrow \infty$.  This limiting value of the Van Hove correlation function is termed the ``ergodicity parameter'' or the ``Edwards-Anderson order parameter'' \cite{Kob-Binder}.  

There are theories, most notably mode coupling theory (MCT)~\cite{MCT1,MCT2,MCT3}, that predict ergodicity breaking to occur at a non-trivial critical packing fraction or temperature as a consequence of non-linear feedback of large fluctuations in equilibrium dynamics.  This dynamic singularity is not supported by experiment or simulation.  Nevertheless, the location of critical packing fractions or densities and temperatures predicted from MCT serve as reasonable estimates for the boundaries between simple liquid behavior and fluctuation dominated dynamics \cite{MCT-onset,MCT-reentrant}, and the approach to that boundary from the simple-liquid side is well described by MCT~\cite{MCT1,MCT2,MCT3}.  It is a regime over which relaxation times change by several orders of magnitude, and MCT's successful description of this regime is a remarkable accomplishment of equilibrium molecular theory.

\section{Fluctuations: dynamic heterogeneity, facilitation and excitation lines}

Having discussed equilibrium mean values of order parameters, we now focus on fluctuations.  This is the subject of dynamic heterogeneity.  Spatial structures associated with fluctuations are resolved by relating parameters fields.  For $K$ and $Q$ of Eqs. (\ref{Activity_K}) and (\ref{Overlap_Q}), these are fields in space and time.  For example,

\begin{equation}
\label{Field1}
K=\int_{V} d\mathbf{r} \,\Delta t \sum_{t=0}^{t_{\mathrm{obs}}} {\,\kappa \left(\mathbf{r},t; \Delta t \right)},
\end{equation}
where integration extends over the net volume of the system, $V$, 
\begin{equation}
\label{Field2}
\kappa(\mathbf{r},t; \Delta t)=\sum_{i=1}^{N} {\left[\mathbf{r}_i(t+\Delta t)-\mathbf{r}_i(t)\right]^2\,\delta \left(\mathbf{r} -\mathbf{r}_i(t) \right)}\,,
\end{equation}
and $\delta(\mathbf{r})$ is Dirac's delta-function.  $\kappa(\mathbf{r}, t; \Delta t)$ is an example of what we call a ``mobility'' field.  Its value at the space-time point $(\mathbf{r},t)$ depends upon whether a particle is found at that point, and further whether that particle moves during the time period $t$ to $t+\Delta t$.

Figure~\ref{fig:heterogeneity} illustrates this specific mobility field for different values of $\Delta t$ in the dense equilibrium WCA mixture.  (The initiation time $t$ is unimportant because for large enough equilibrium systems, such as those pictured, any $t$ is equivalent to any other $t$. In other words, equilibrium has time-translation symmetry.)  Two temperatures are considered, one above the onset to glassy dynamics, and one below the onset.  The most striking feature is the differing degrees of heterogeneity for the two cases.  Panels at the same value of $\Delta t / \tau_e$ illustrate fields at the same relative stages of average structural relaxation.  The differing heterogeneities are thus not simply reflections of differing mean fields.

\fig{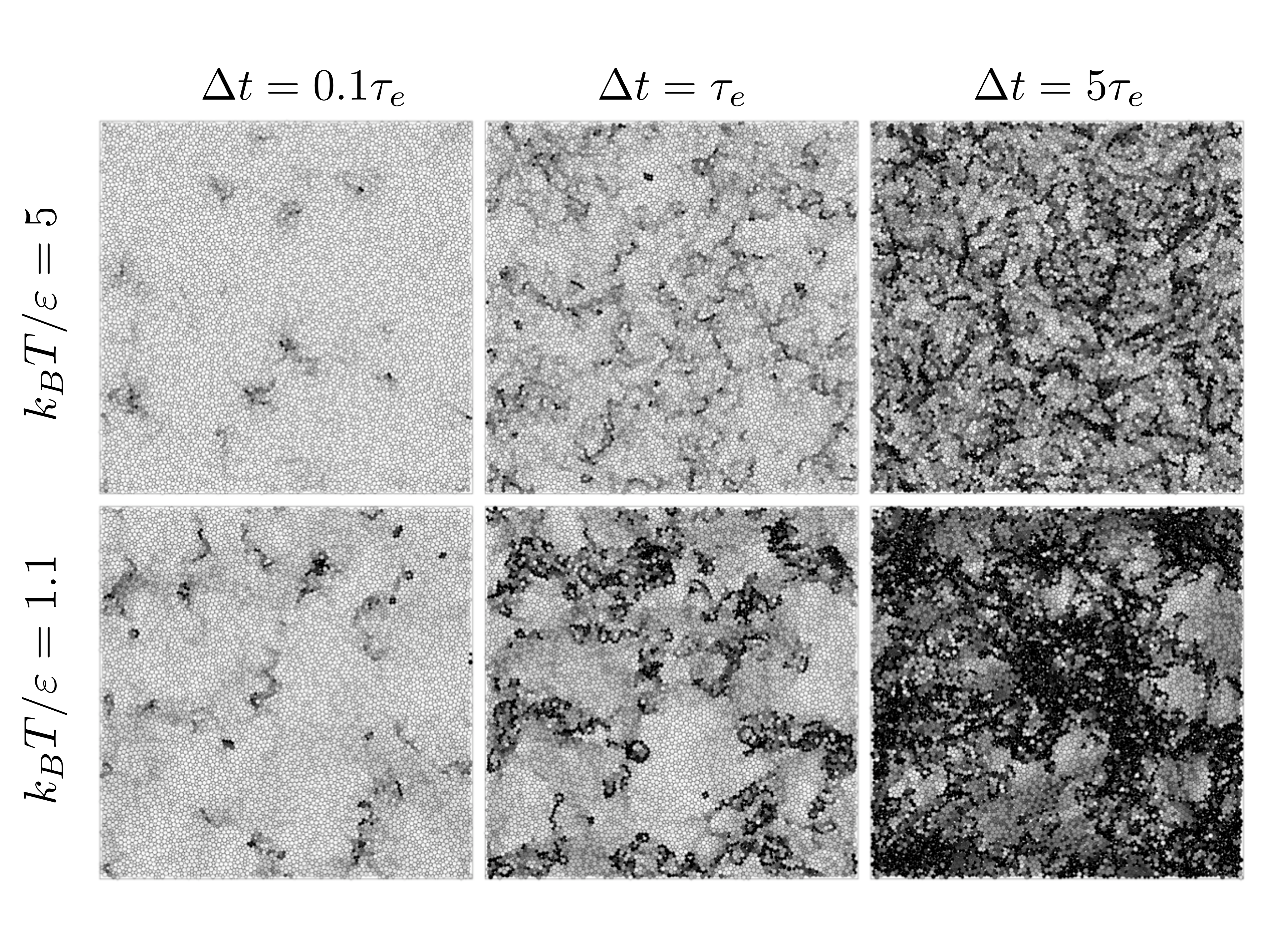}{Dynamic heterogeneity for the $d=2$ WCA mixture considered in Fig.~\ref{fig:Corrfuncs}.  The pictures are renderings of the mobility field $\kappa(\mathbf{r},0;\Delta t)$ for typical equilibrium trajectories of $10^4$ particles.  The displacement time, $\Delta t$ is given in reference to the $1/e$-time for the Van Hove self correlation functions at the corresponding temperatures. The rendering shades particles according to the size of the particle's displacement from its initial position.  If the $i$th particle's displacement is nil, i.e.,$|\mathbf{r}_i(\Delta t)-\mathbf{r}_i(0)|=0$, the particle is pictured as white (excepting a thin circle indicating its diameter).  As the particle displacement grows, the particle acquires an increasing shade of grey, becoming completely black when $|\mathbf{r}_i(\Delta t)-\mathbf{r}_i(0)|\geqslant \sigma$. }{fig:heterogeneity}{ht}{.8}

The darker the particles the more they have moved from their initial positions.   After a long enough time $\Delta t$, all particles will move, and the pictures for both high temperature and low temperature cases will be uniformally black.  But on the way to getting there, the low temperature mobility field is structured with thick connected mobile domains separating large immobile domains.  The latter are cross sections of what we call ``bubbles in space-time,'' about which we will soon say more.  For now, notice that these large voids of immobility are present over extended periods of time in the low temperature case, whereas they are absent for all but the shortest $\Delta t$ in the high temperature case.  This juxtaposition indicates that spatial correlation of dynamics in the low temperature trajectory is much more significant than it is in the high temperature trajectory. 

``Large'' is used here to describe lengths that are large compared to typical equilibrium correlation lengths, such as the coarse graining length over which the pair correlation function becomes structureless.  This bulk correlation length is roughly $2\pi/k_{\mathrm{min}}$, where $k_{\mathrm{min}}$ is the smallest wave-vector where the structure factor first deviates significantly from its limiting small wave-vector value. For dense fluids (or glasses), this length is usually not larger than one or two diameters of the principal molecular species.  In the high temperature system, one may infer from Fig.~\ref{fig:heterogeneity} that mobility coarse grained on this length scale spreads uniformly throughout the system so that its behavior is well approximated by the mean mobility.  In contrast, in the low temperature system, motion does not spread so uniformly, and the typical mobility coarse grained over the bulk correlation length is not well approximated by its mean.  In other words, Fig.~\ref{fig:heterogeneity} demonstrates that dynamics of the high temperature system is like that of the mean field, while dynamics of the low temperature system is fluctuation dominated.

Similar pictures of dynamic heterogeneity can be produced using any number of fields that capture local motion,  for example, with the real part of ${\cal Q}(\mathbf{r},t;k,\Delta t)$ with $k\approx 2\pi/\sigma$.  This field is related to the net overlap $Q$ in the same way that $\kappa(\mathbf{r},t;\Delta t)$ is related to the net mobility $K$.  Fluctuations in this field correspond to fluctuations in the self-correlation function, and their structure factors, $\chi_4(q; k,\Delta t) = \int d \mathbf{r'} \langle {\cal Q}(\mathbf{r},t;k,\Delta t)\,{\cal Q}(\mathbf{r+r'},t;k,\Delta t)\rangle e^{i\mathbf{q} \cdot \mathbf{r'}}$, have been studied with computer simulations and theory by several workers~\cite{Lacevic,Whitelam,Toninelli,chi4-bb1,chi4,Szamel-Flenner,chi4-bb2,chi4-bb3,Stein}.  $\chi_4(q; k,\Delta t)$ provides a useful quantitative measure of dynamic heterogeneity in an equilibrium system.  It does not, however, provide a direct measure of space-time structure for transitions that break time-translational symmetry.  For that case, the signature of a transition would be the divergence of mean square fluctuations of extensive order parameters like $K$ and $Q$, a divergence that grows super-extensively in $t_{\mathrm{obs}}$.  From Eqs.~(\ref{Activity_K}) and (\ref{Overlap_Q}) we see that this signature can result from correlations between fields like ${\cal Q}(\mathbf{r},t;k,\Delta t)$ and ${\cal Q}(\mathbf{r'},t';k,\Delta t)$ as $|t-t'| \rightarrow \infty$.  The function $\chi_4(q; k,\Delta t)$ contains information about $t=t'$ only.

\subsection{Facilitated and hierarchical dynamics}

It is informative to enlarge Fig.~\ref{fig:heterogeneity} to the point where one can easily visualize the great extent to which particle are packed tightly.  With that view in mind, it is interesting to know what is found from a movie of the mobility field with $\Delta t$ growing continuously.  Unedited, viewers are struck by high frequency motions, with particles flipping rapidly back and forth between light and dark shades.  These distracting high frequency motions can be filtered out by replacing $\mathbf{r}_i(t)$ and $\mathbf{r}_i(t+\Delta t)$ in Eq.\ (\ref{Field2}) with $\bar{\mathbf{r}}_i(t)$ and $\bar{\mathbf{r}}_i(t+\Delta t)$, respectively, where $\bar{\mathbf{r}}_i(t)=(1/\delta t)\int_{0}^{\delta t} dt \,\mathbf{r}_i(t)$. Rendering the resulting time-coarse grained mobility field, $\bar{\kappa} (\mathbf{r},t;\Delta t)$ with $\delta t\approx0.01\tau_e$, proves illustrative.  Movies created in this way \cite{Movies} show the system possessed by low frequency low amplitude motions.  These are the soft modes of a disordered solid.  Darkening of particles signifying significant particle displacements takes place on time scales large compared to the apparent periods of those small amplitude motions, and this darkening takes place in the form of surges.  String-like structures tens of molecules long emerge and then retract from central dark regions. The frequent retractions are indicative of the degree to which motion is restricted.  An unconstrained random walk in two-dimensions would rarely lead to a perfect reversal over several atomic lengths.  

These surging strings are motions catalogued by Glotzer and her co-workers~\cite{Glotzer-review}.   In a small system, these would be the only structures apparent before the entire system became dark.  But for the size system pictured in Fig.~\ref{fig:heterogeneity}, much more is seen.  In particular, after several thrusts, a string emitted from one dark region no longer retracts, but rather blossoms into a larger fixed dark region that melds with the original dark region.  From the new and now larger dark region, further strings surge, and the story repeats again at a larger length scale.  These features exemplify a type of dynamics that is called ``facilitated'' and ``hierarchical.''  It is facilitated because mobility in a region of space leads to motion (or relaxation) in an adjacent region of space.  It is hierarchical because dynamics on smaller length scales (surging) is more frequent than motion on longer length scales (sticking and growing after several thrusts).  Provided temperature is lower than the onset temperature, the great majority of events observed in renderings of $\bar{\kappa} (\mathbf{r},t;\Delta t)$ follow this type of path -- facilitated and hierarchical dynamics.

\fig{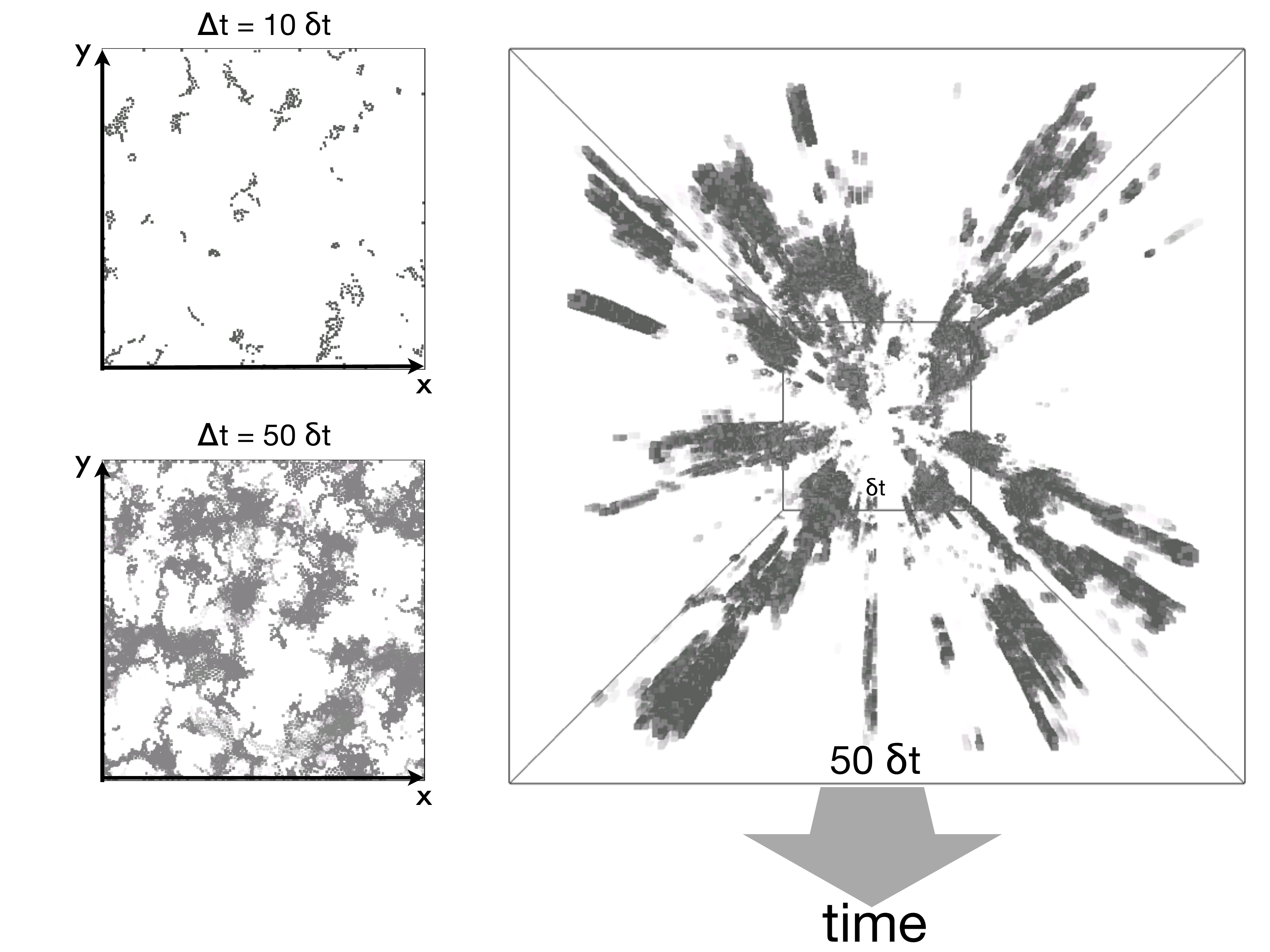}{Excitation lines viewed from a bubble in space-time for the low temperature WCA mixture depicted in Fig.\ \ref{fig:heterogeneity}.  Upper left panel shows the mobility field for the darkest (most mobile) particles when $\Delta t = 10 \delta t = 0.1 \tau_e$.  These dark regions are the elementary mobility excitations of the model.  The right panel shows the trajectory of these excitations over a time span of $t= 50\delta t$.  The bottom left panel shows the mobility field when $\Delta t = 50\delta t$ which can be thought of as the accumulation of the mobilities from 100 times slices of the trajectory shown in the right frame. }{fig:ExcitationLines}{ht}{.8}

\subsection{Excitation lines, decoupling and continuous-time random walks}

Another way to think about dynamical heterogeneity and facilitation is with the concept of excitation lines~\cite{Garrahan-Chandler-2002}.  To the extent that dynamics is facilitated, excitations should form lines in space-time.  This is true because facilitated dynamics requires an adjacent excitation for the birth of an excitation, and from time-reversal symmetry, an adjacent excitation is also required for the death of an excitation.  Hence, space-time in a system with facilitated dynamics is structured with strings of excitations that are directed in time, and fluctuations in numbers of excitations come from these lines dividing and coalescing.  At any time slice, there is an equilibrium distribution of these excitations, and at equilibrium, this distribution is preserved from one time slice to the next.  The dynamic heterogeneity of Fig.~\ref{fig:heterogeneity} should then emerge from overlaying the time series of these distributions.

Figure~\ref{fig:ExcitationLines} exhibits this type of structure. In this case, excitations are local regions of mobility, specifically the darkened regions located by the mobility field in upper-left panel of Fig. \ref{fig:ExcitationLines}.  These excitations are identified from the system's dynamics for the microscopic time $\delta t$.  For a glass forming liquid, they are sparse, like an ideal gas.  When these dilute excitations at time $t$ are added to those at times $t+\delta t$, and those at time $t+2\delta t$, and so on up to those at $t+n\delta t$, the resulting field is approximately the more expansive dynamic heterogeneity field shown in the lower left panel of Fig. \ref{fig:ExcitationLines}, namely $\kappa(\mathbf{r}, t, \Delta t)$, where $n \delta t = \Delta t$.  But one can also separately view each time slice of the trajectory, $\kappa(\mathbf{r}, t, \delta t)$ for $t = \delta t, 2\delta t, ..., n\delta t$.  This alternative produces the excitation lines of the super-cooled WCA mixture shown in the right panel of Fig. \ref{fig:ExcitationLines}.  On viewing it we can see that indeed the space-time structure this atomistic model of a glass forming liquid is as imagined from the perspective of facilitated dynamics.

\fig{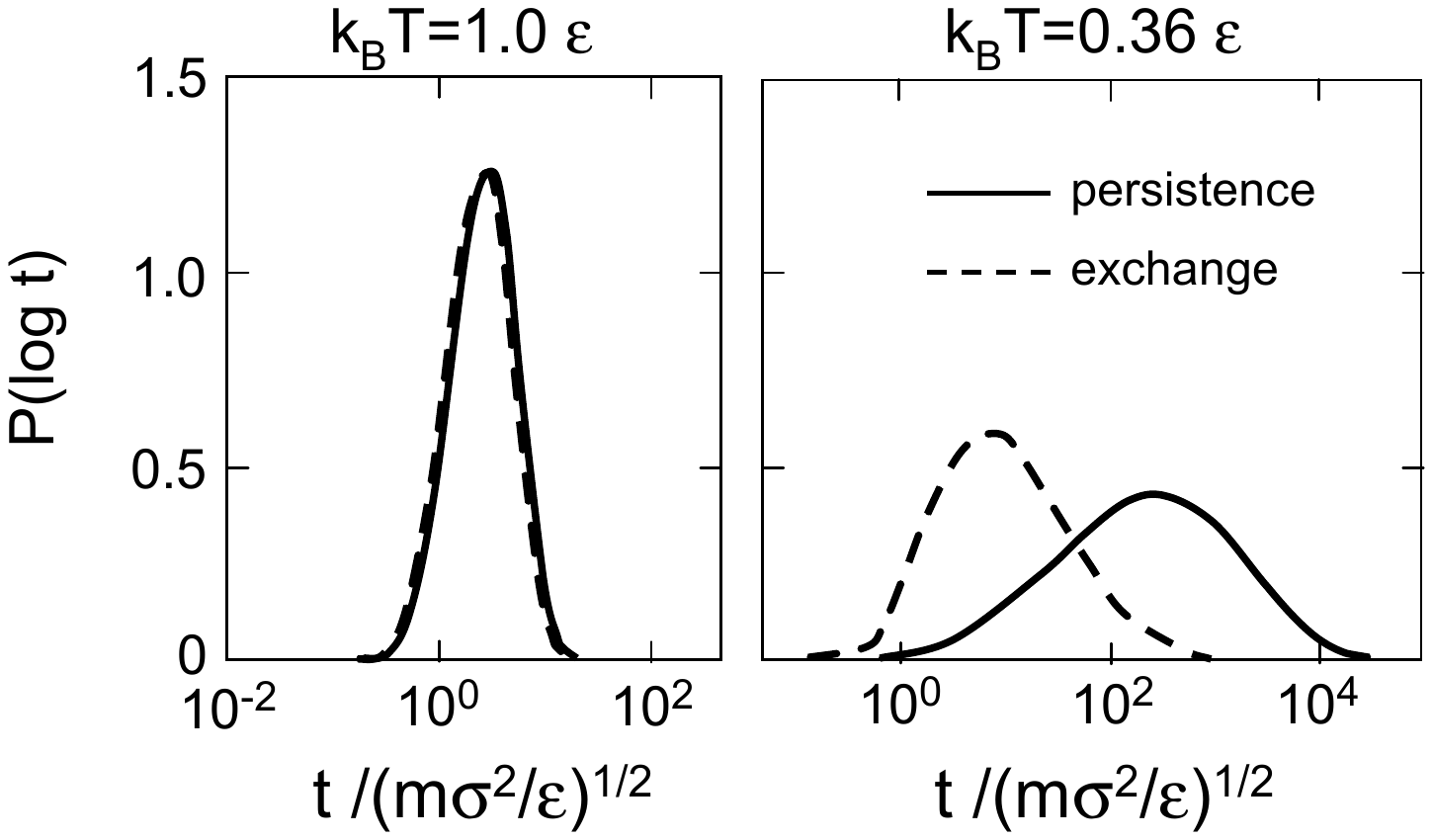}{Probability distributions of log exchange time and log persistence time for the $d=3$ WCA mixture for events defined as particles displacement of size $a = 0.5 \sigma$.  Adapted from Ref.\ \cite{wcadecoupling} where system details are described.   The onset temperature for this system is $T_{\mathrm{o}}= 0.6 \epsilon / k_{\mathrm{B}}$ \cite{Elmatad}.  As such, the left panel refers to a normal liquid temperature, and the right panel refers to a super-cooled liquid.}{fig:ExVsPer}{ht}{.8}

We first described this structure in Ref.\ \cite{Garrahan-Chandler-2002}, where we showed how its geometry provides
explanations for quantitative aspects of dynamic heterogeneity. In Ref.\ \cite{Garrahan-Chandler-2002}, and in several subsequent papers, we have used kinetically constrained lattice models
(KCMs), models where facilitation
is a given or presumed property of dynamics.  From Fig.\ \ref{fig:ExcitationLines}, we see this property
emerges from Newtonian dynamics of a sufficiently supercooled or compressed material.
A few published papers have focused on how facilitation can be an emergent property~ \cite{plaquette1,plaquette2,Vogel,Chamon,Kennett,Dauchot2}.
The surging events described in the previous subsection coincide with the geometry of excitations in the KCM known as the ``East'' model \cite{Jackle-Eisinger};   see, in particular, Figs. 2, 3 and 5 of Ref.\ \cite{Garrahan-Chandler-2002}. It is therefore perhaps not surprising that the temperature dependence for relaxation times predicted by this simple model and its generalizations to higher dimensions \cite{KCMs} is the temperature dependence shown to agree so well with experiment in Fig.\ \ref{fig:tExp}.  

One consequence of the excitation-line structure of space-time is that dynamical processes can be partitioned according to whether they coincide with exchange events or persistence events.  The distinctions between these two classes of processes are responsible for a host of non-linear phenomena that are characteristic of deeply super-cooled liquids.  To understand, consider a tagged molecule in a glass forming material.  It might be a probe molecule that is investigated with single molecule spectroscopy, or it might be one of the molecules comprising the liquid itself.  In view of Fig.\ \ref{fig:ExcitationLines}, we see that over a time frame $\delta t$ it is most likely that this particle is immobile.  In other words, it most likely sits in a void of mobility -- what we call a bubble in space-time.  Nevertheless, after some period of time, an excitation line will intersect an immobile particle, whereupon this particle will be able to move.  It will continue to move until it steps out of the excitation line, or the line passes it by, whereupon the tagged particle will again be immobile.  The picture we draw, therefore, is a dynamics where a tagged particle changes intermittently between mobile and immobile states \cite{Review-Ediger,Weeks-et-al}.

The distributions of times characterizing this intermittent behavior can be viewed in terms of the time from its initial condition before it moves, and the time between two moves.  The first of these is called a ``persistence'' time, $t_{\mathrm{p}}$, and the latter is called an ``exchange'' time,  $t_{\mathrm{x}}$ \cite{YJ2}.  (In the literature on renewal processes, exchange and persistence times are sometimes called ``waiting'' and ``excess life'' times, respectively~\cite{Grimmett}.)  For uncorrelated sequences of events, persistence times and exchange times have the same distribution, and the probability that an event has not occurred in a time $t$ is $\exp(-t/\tau)$, where $\tau = \langle t_{\mathrm{x}} \rangle = \langle t_{\mathrm{p}} \rangle$.  Such intermittencies are known as ``Poisson'' processes.  In the case of glassy dynamics, however, event statistics is far from Poissonian because an event becomes more likely when a similar event has occurred more recently.  In other words, facilitation implies that events are bunched in time.  As a result, with the correlated dynamics of a super-cooled glass former, exchange times are typically much shorter than persistence times.  Figure \ref{fig:ExVsPer} demonstrates this behavior for the distributions of persistence and exchange times in a $d=3$ dimensional WCA mixture.  At normal liquid conditions the two distributions coincide and are Poissonian.  At supercooled conditions, however, the two distributions decouple with $\langle t_{\mathrm{p}} \rangle \gg \langle t_{\mathrm{x}} \rangle$.

One experimental manifestation of decoupling is the breakdown of Stokes-Einstein relations like $D \eta /T \approx$ constant or $D \,\tau_{\alpha} \approx$ another constant.  Here, $D$ stands for a translational diffusion constant and $\eta$ stands for shear viscosity.  Because the average persistence time is the longer of the two mean relaxation times, it is an estimate of the structural relaxation time, $\tau_{\alpha} \approx \langle t_{\mathrm{p}} \rangle$.  On the other hand, the time between random walk steps is $ t_{\mathrm{x}} $ so that $D \approx a^2/\langle t_{\mathrm{x}} \rangle$, where $a$ is a step size.  As such, $D \, \tau_{\alpha} \approx a^2 \langle t_{\mathrm{p}}\rangle / \langle t_{\mathrm{x}} \rangle$.  The Stokes-Einstein relations hold when $\langle t_{\mathrm{p}}\rangle / \langle t_{\mathrm{x}} \rangle \approx 1$.  This is the mean field result, and it is obeyed to a good approximation at standard liquid conditions.  But in the fluctuation dominated regime of super-cooled liquids, the ratio can grow.  The result~\cite{YJ1} is a large failure of the Stokes-Einstein form that is seen experimentally~\cite{Chang-Sillescu,Swallen,Schweizer-review}, and is illustrated for an atomistic simulation in Fig.\ \ref{fig:decoupling}.

Typical excitation voids or bubbles of space-time persist for the time scale of structural relaxation, while particles diffuse by, in effect, surfing on the excitation lines that surround these voids.  It is this surfing that makes $D$ so much larger than anticipated by the Stokes-Einstein relationship.  On the other hand, by applying a large enough external force to a tagged particle, the particle can be pulled out of an excitation line.  In this way, a particle's drift velocity can actually decrease with increasing force.  This negative response is yet another hallmark of fluctuation dominated correlated dynamics that has been observed in simulations.  See for instance, Refs.\cite{Sellitto1, DKnegative}. 

In cases of non-hierarchical dynamics (i.e., strong glass formers), fluctuation effects become irrelevant in dimensions $d\geqslant2$ \cite{Rob}.  For these systems, $\langle t_{\mathrm{p}}\rangle / \langle t_{\mathrm{x}} \rangle$ will not grow upon lowering $T$.  As a result, the predicted~\cite{pnas} crossover from fragile behavior to strong behavior upon lowering temperature might be tested by monitoring the behavior of $D \tau_{\alpha}$.  This possibility has been examined theoretically with KCMs~\cite{Pan1, Pan2}, and the trends established with those models seem to be observed experimentally~\cite{Chang-Sillescu,ChenSE}.

There is a stochastic formalism that codifies our picture of the dynamics of a tagged molecule in terms of a series of periods of mobility or activity, with displacements larger than uninteresting small amplitude vibrations, punctuated by large quiescent periods of immobility or inactivity.  It is the theory of continuous-time random walks (CTRW) \cite{Montroll}.   The formalism provides a means to distinguish the first from subsequent steps in a particle's trajectory.  In the context of glassy dynamics, the first step coincides with persistence.  It is the period before the tagged particle is intersected by an excitation line.  The subsequent steps coincide with exchange events.

\fig{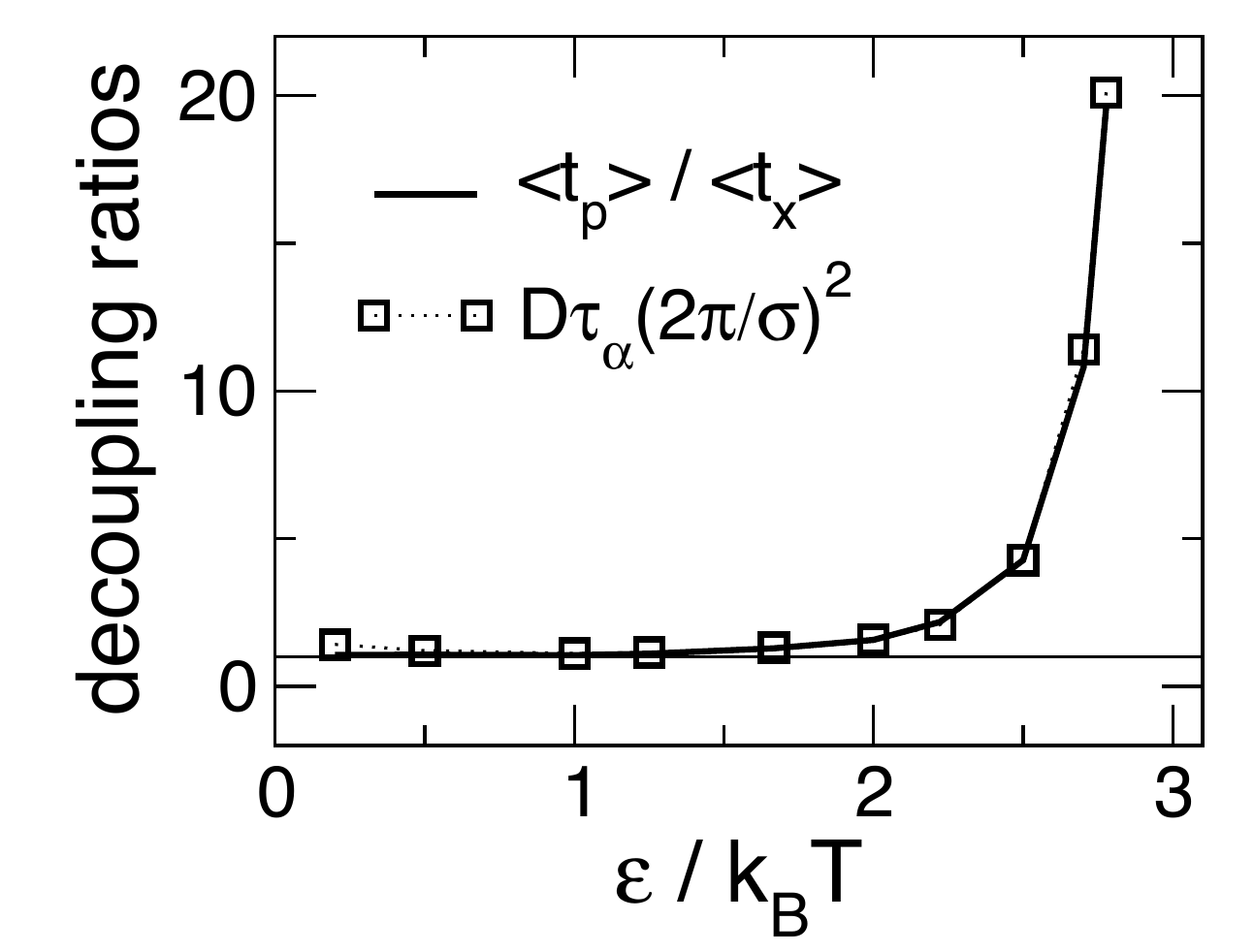}{Decoupling ratios for the $d=3$ WCA mixture of Fig.\ \ref{fig:ExVsPer}.  Adapted from Ref.\ \cite{wcadecoupling}.   We plot the quantity $\langle t_{\mathrm{p}} \rangle / \langle t_{\mathrm{x}} \rangle$ (dark line) which measures deviations from Poissonian statistics a function of temperature.  At high temperatures the two average timescales are the same and this quantity is unity.  At low temperatures the persistence time rapidly decouples from the exchange time.  We also plot the deviation from the Stokes-Einstein relation $D k^2 \tau_\alpha$, for $k = 2\pi/\sigma$.  The decoupling between self-diffusion and alpha-relaxation timescale coincides with that of persistence and exchange.  The onset temperature for this liquid mixture as it is defined by Eq.\ref{TauFragile}, $T_{\mathrm{o}}= 0.6 \epsilon / k_{\mathrm{B}}$ \cite{Elmatad}, is in harmony with the onset temperature for decoupling.}{fig:decoupling}{ht}{.45}

The CTRW formalism has been applied to the Van Hove self correlation function with this perspective~\cite{EPL}, and it yields a result that is particularly vivid in the regime where $\langle t_{\mathrm{p}}\rangle \gg \langle t_{\mathrm{x}}\rangle$:
\begin{equation}
F(\mathbf{k},t) \approx P(t) + [1-P(t)] \exp{(-D k^2 t)} .
\label{Fapprox}
\end{equation}
Here, $P(t)$ is the probability that a particle persists in its initial configuration for a time $t$, and $D$ is the translational self-diffusion constant.  The first term on the right-hand-side of (\ref{Fapprox}) is the persistence contribution to the mean concentration of overlap with initial conditions, the second term is the diffusive contribution. 
The balance between the two terms in (\ref{Fapprox}) depends on the value of the wavevector $k = | \mathbf{k} |$.  That is, the relaxation timescale of self-correlations is lengthscale dependent \cite{EPL} 
\begin{equation}
\tau_e(k) \approx \langle t_{\mathrm{p}}\rangle + \frac{1}{k^2 D} \approx \langle t_{\mathrm{p}}\rangle + \frac{\langle t_{\mathrm{x}}\rangle}{a^2 k^2} .
\label{tauk}
\end{equation}
For large wavevectors, such as $k=2\pi/\sigma$ in the WCA mixture, the first terms in (\ref{Fapprox}) and (\ref{tauk}) dominate, so that $\tau_e(k) \approx \tau_\alpha \approx \langle t_{\mathrm{p}}\rangle$.  For small enough wavevector the second terms in (\ref{Fapprox}) and (\ref{tauk}) dominate, and relaxation becomes exponential with the diffusive (i.e., Fickian) relaxation time $\tau_e(k) = 1/Dk^2$.  Comparing terms of Eq.\ (\ref{tauk}), we identify the characteristic lengthscale for the crossover between non-Fickian to Fickian regimes: $l_{\mathrm{F}} \equiv \sqrt{ D \tau_\alpha} \propto \sqrt{\langle t_{\mathrm{p}}\rangle/\langle t_{\mathrm{x}}\rangle}$.
The lengthscale dependent timescales and the non-Fickian/Fickian crossover are central predictions arrived at from KCMs and facilitation theory, predictions that are in harmony with both experiments \cite{Ediger-recent} and simulations \cite{BerthierPRE,Szamel-fickian}.   The physical picture combined with the CTRW formalism has also has been used to analyze the $\chi_4$-function~\cite{chi4}, to interpret negative response~\cite{DKnegative}, and to explain exponential tails \cite{Chaudhuri} observed in real-space self-correlation function \cite{Stariolo}.  It has also been extended to the study of ``metabasin'' transitions \cite{Heuer}, making a connection between KCMs and the energy landscape perspective \cite{Review-Debenedetti}.

\section{Order-disorder in space-time}

The previous section described consequences of dynamic heterogeneity in structural glass forming liquids -- broad distributions of length scales, time scales, decoupling, and negative response.  Explicit pictures of dynamic heterogeneity, Figs.\ \ref{fig:heterogeneity}, look
much like those that one might see by studying the emergence or coarsening of an equilibrated
state from a metastable phase \cite{OnukiBook}.  It is as if there is an initial metastable immobile phase from which the stable ergodic or equilibrated phase emerges.  In this section we discuss how indeed dynamic heterogeneity is the precursor to a phase-transition in trajectory space \cite{Garrahan-Chandler-2002,Mauro}.  Doing so uses the so-called method of ``large deviations'' in generalizing traditional equilibrium statistical mechanics to a statistical mechanics of trajectory space~\cite{Lebowitz,Lecomte,Touchette}.

We use the symbol $x_t = (\mathbf{r}_1(t),\mathbf{r}_2(t),...,\mathbf{r}_i(t),...)$ to denote the point in phase space for the system at time $t$, and denote a trajectory with $x(t) = (x_0, x_{\delta t}, x_{2 \delta t},..., x_t,..., x_{t_{\mathrm{obs}}})$, where $\delta t$ is the time step for the trajectory, and the trajectory runs for a total of $t_{\mathrm{obs}}/\delta t$ time steps.  For any system, there is an ensemble of possible trajectories.  We use $P[x(t)]$ to stand for the distribution of trajectories at conditions of equilibrium.  For deterministic trajectories, this distribution would be the equilibrium distribution of initial conditions times a product of delta-functions, one for each time step, specifying the rule by which the phase space point at time $t$ is reached from that at time $t-\delta t$.  For stochastic trajectories, $P[x(t)]$ would contain a product of transition probabilities more general than delta-functions.  (We are assuming dynamics is Markovian.)  The specific form of $P[x(t)]$ is not important.  All that is important is that $P[x(t)]$ is normalized, it preserves an equilibrium distribution, and it is time-reversal symmetric.

The phenomenon we describe with this notation emerges from non-equilibrium dynamics -- a non-ergodic glass phase that becomes the stable phase in an ensemble that is driven away from equilibrium.  A way to construct non-equilibrium ensembles is to bias the equilibrium ensemble according to the value of some order parameter.  For example, we can use the order parameter $K[x(t)]$ given in Eq.\ (\ref{Activity_K}) and change $P[x(t)]$ to
\begin{equation}
\label{s-ensemble}
P_s[x(t)] \propto P[x(t)]\exp \left( -sK[x(t)] \right),
\end{equation}
with the normalization constant $1/Z_s$, where $Z_s = \langle \exp \left( -sK[x(t)] \right) \rangle$. In analogy with equilibrium statistical mechanics, the parameter $s$ is to $K[x(t)]$ what $1/k_{\mathrm{B}}T$ is to energy or what pressure times this factor is to volume.

A positive value of $s$ biases the ensemble towards trajectories that show little activity, a negative value of $s$ biases the ensemble towards trajectories that show much activity.  By adjusting the value of $s$ one may push the ensemble far from regions that are most probable in the equilibrium distribution.  In the limit of very large observation times, the quantity $t_{\mathrm{obs}}^{-1} \ln Z_s$ converges to a time-independent function often termed the ``large deviation'' function.  In other words, $\ln Z_s$ is like a free energy of trajectory space, one that is extensive in time (and in space).  A singularity in its density in the thermodynamic limit, $[V^{-1}t_{\mathrm{obs}}^{-1}\,\ln Z_s ]_{V,t_{\mathrm{obs}}\rightarrow \infty}$, such as a discontinuous derivative with respect to $s$, is the signature of a space-time phase transition with the chosen order parameter.  Such behavior has been established to occur in several different KCMs, independent of any equilibrium phase transitions~\cite{Fred, Fred-long}.

The average order parameter in the non-equilibrium ensemble is
\begin{equation}
\label{K_s}
K_s =\,-\,\partial \ln Z_s / \partial s\,\,=\,\, Z_{s}^{-1}\, \langle K[x(t)] \exp \left( -sK[x(t)] \right) \rangle \,\,.
\end{equation} 
In the context of molecular simulation, the second equality of Eq.\ (\ref{K_s}) looks like a familiar expression used when re-weighting states in a thermodynamic perturbation theory calculation~\cite{Chandler, FrenkelSmit}.  The difference, of course, is that here we are averaging over trajectories, not states.  As such, applications of these equations with numerical simulation require simulation techniques that do importance sampling of trajectory space.  Transition path sampling~\cite{TPS} is a general methodology of this type, and it has been used in this way to implement the re-weighting of trajectory space~\cite{Order-disorder, Mauro, Jack}. Figures \ref{fig:sketch} \& \ref{fig:sEnsemble} show results obtained for the Kob-Andersen mixture of Lennard-Jones particles.

The evidence for a first-order phase transition induced by a small but positive value of $s=s^*$ is compelling.  In particular, on increasing system size and trajectory length, the change in $K_s$ on passing from $s<s^*$ to $s>s^*$ becomes more singular, and the order-parameter probability distribution becomes more bi-modal.  Cleary, at $s=s^*$ the system exhibits the signatures of macroscopic phase coexistence -- a coexistence between an ergodic phase and a non-ergodic phase.  At $s=s^*$, the material is not simply micro-heterogeneous, as it is at equilibrium.  Rather here, pushed out of equilibrium, the domains of inactivity are as large as the total system.  In terms of the equilibrium pair distribution function, $g(r)$, the structures of the equilibrium phase and the non-equilibrium phase are indistinguishable.  Yet as measured by the van Hove correlation function, the non-equilbirium phase is non-ergodic -- there is finite overlap with initial conditions for all times.  

For the KCMs that have been studied, the non-equilibrium active-inactive transition takes place at $s^*=0^+$~\cite{Fred,Fred-long}, which means that equilibrium dynamics of these systems takes place at coexistence with the inactive dynamical phase.  This location reflects that kinetic constraints in these KCMs are hard, i.e., they cannot be violated.   For super-cooled liquids in general, kinetic constrains can be overcome, but in a way that becomes progressively unlikely as temperature decreases.  One would therefore expect the coexistence point $s^*$ to be at a non-zero value, i.e. at $s=0$ the inactive phase is destabilized by constraint violating rearrangements.  Nonetheless, at super-cooled conditions we expect $s^*$ to be small, so that the equilibrium dynamics lies close to coexistence with the non-ergodic phase.

\fig{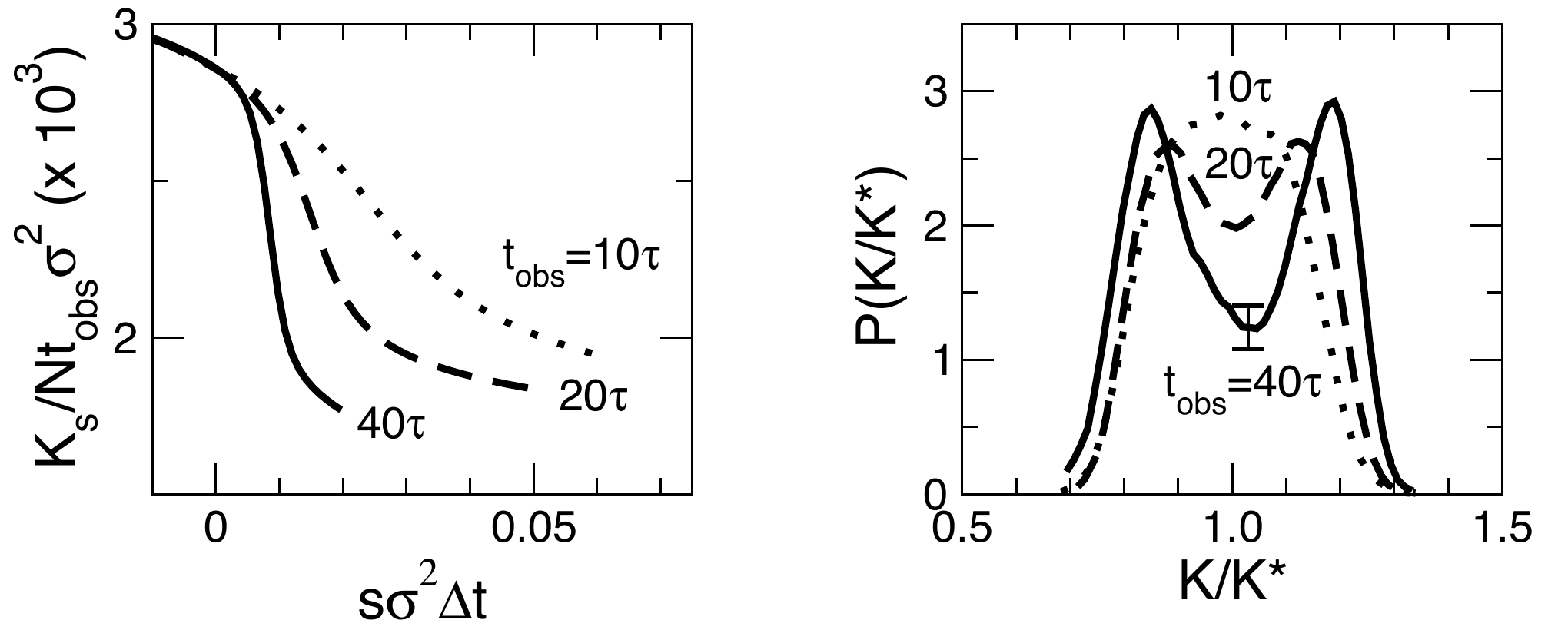}{Evidence for first-order phase transition in space-time between an ergodic phase and a disordered non-ergodic phase.  Average space-time order parameter  $K_s$ as a function of biasing field $s$, from molecular dynamics simulations of the Kob-Andersen Lennard-Jones liquid for $N=150$ total particles and reduced density $(N_A \sigma_A^3 + N_B \sigma_B^3)/V=1.05$, at super-cooled conditions, $T= 0.6\epsilon / k_{\mathrm{B}}$.  The onset temperature as defined in Eq.\ (\ref{TauFragile}) for this system is $T_{\mathrm{o}}= 0.8\epsilon / k_{\mathrm{B}}$ \cite{Elmatad}.  The field $s$ couples to the net mobility $K$.  As the length of trajectories increases, the crossover in $K_s$ becomes sharper and happens at smaller values of $s$.  For large $t_{\rm obs}$, the order parameter distribution at coexistence, $s=s^*$, becomes bimodal, as expected for a first-order transition.  While discriminating different regions of trajectory space with different values of $s$, the sampling avoids regions of trajectory space with ordered non-ergodic crystals by biasing against crystal order parameters.  Adapted from Ref.\ \cite{Order-disorder}, where system size and other simulation details are described.}{fig:sketch}{ht}{.7}

\fig{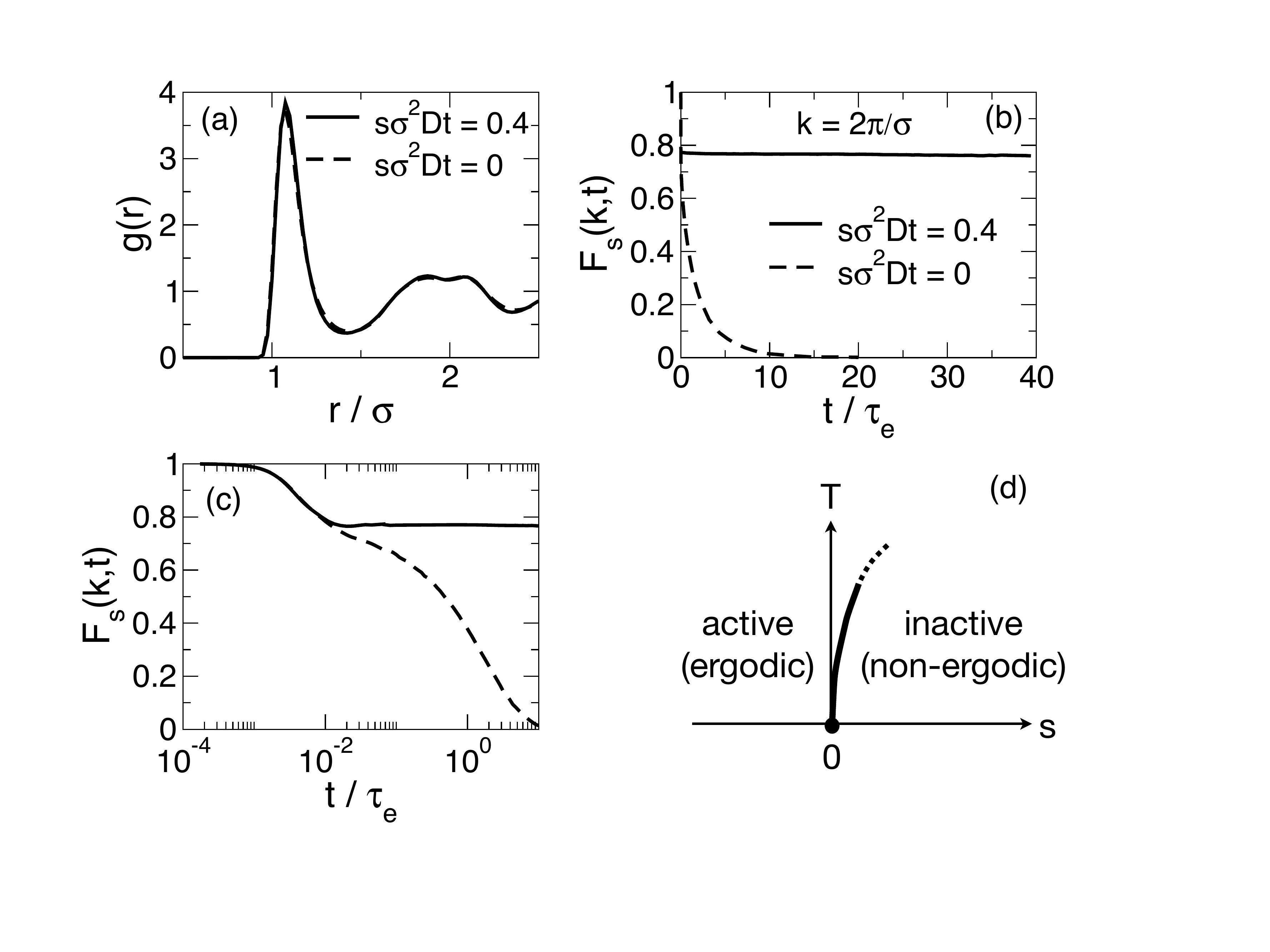}{Structural and dynamical measures of ergodic and non-ergodic phases of the $d=3$ Lennard-Jones mixture of Kob and Andersen, for the same parameters as in Fig.\ \ref{fig:sketch}.  The field $s$ couples to the net mobility $K$.  (a) The radial distribution functions for the equilibrium ergodic phase, $s=0$, and non-equilbrium non-ergodic phase, $s>0$.  (b) The Van Hove self correlation functions for the ergodic and non-ergodic phases, illustrated over a range of times very large compared to the equilibrium structural relaxation time $\tau_e$.  (c) The Van Hove self correlation function, illustrated over a range of the order of the equilibrium structural relaxation time.  (d)  Qualitative phase diagram depicting the ergodic-nonergodic coexistence line in the $s-T$ plane.  Adapted from Ref.\ \cite{Order-disorder}, where system size and other simulation details are described.  }{fig:sEnsemble}{ht}{.7}

Figures \ref{fig:sketch} \& \ref{fig:sEnsemble} show that the inactive states exist at temperatures below those of the onset of super-cooled behaviour.  If one places the system in a configuration of the inactive phase, the system will remain inactive for times much longer than the typical structural relaxation time, even without biasing the dynamics.  To the extent that $s^*$ goes to zero in the thermodynamic limit, the inactive phase can be called an ``ideal glass".  Its micro-states are vanishingly rare in the equilibrium ensemble, but once prepared, molecular dynamics will produce configurations that always overlap with that state.  In other words, initial conditions are space-time surfaces which can bias the dynamics towards the active or inactive phase.  But initial conditions taken at random from the equilibrium phase will essentially always select this equilibrium phase -- the active dynamical phase~\cite{Jack}. 

To make this point explicit, imagine partitioning $Z_s$ according to its initial conditions, 
\begin{equation}
\label{partition}
Z_s = \int dx_0 \,\rho(x_0)\, Z_s(x_0)\,=\,\langle Z_s(x_0)\rangle,
\end{equation}
where $\rho(x_0)$ is the equilibrium distribution of initial states $x_0$.  As we have noted, in an ergodic phase, the net partition function, $Z_s$ is exponentially in trajectory length $t_{\mathrm{obs}}$.  The same is true for $Z_s(x_0)$.  For large enough $t_{\mathrm{obs}}$, the annealed average over initial conditions, Eq.\ (\ref{partition}), is therefore dominated by those initial states with the largest values of $(1/t_{\mathrm{obs}})\ln Z_s(x_0)$.  In this way, the non-ergodic phase can dominate even when its equilibrium weight is negligible.  On the other hand, if the non-ergodic states have negligible equilibrium weight, the partition function for a quenched average over initial conditions,
\begin{equation}
\label{quenched}
Z_s^{(\mathrm{q})} = \exp[\langle \ln Z_s(x_0)\rangle],
\label{Zq}
\end{equation}
will always yield the partition function for the ergodic trajectories.  In other words, the inactive phase can only be reached by some form of non-equilibrium driving \cite{Complexity}. 

This difference between quenched and annealed averages is a hallmark of rare region statistics \cite{Vojta}.  Due to the proximity of bulk phase coexistence with the inactive phase, equilibrium trajectories display pronounced space-time bubbles of inactivity, and as demonstrated with KCMs, the largest or rarest of these inactive regions give rise to super-Arrhenius relaxation times and stretched time correlations \cite{SollichEvans,Garrahan-Chandler-2002}.   Singular behaviors driven by surfaces imply surface phase-transitions \cite{Wetting}.  For the transition between ergodic and non-ergodic phases, the ``surface'' is the initial conditions in space-time, and ``wetting'' is the overlap with initial conditions.  From this perspective, the two-step correlators of Fig.\ \ref{fig:Corrfuncs} can be viewed as a pre-wetting profile, and the broken time-symmetry illustrated in Fig. \ref{fig:sEnsemble} can be viewed as a wetted profile.

\section{Summing up}

The order-disorder phenomena discussed in the previous section is a bona fide phase transition between an ergodic state and non-ergodic state, with a partition function that is singular in a thermodynamic limit.  It is, however, a phase transition that occurs away from equilibrium, with a partition function of trajectory space.  With this transition in mind, it becomes comprehensible why a glass can form and why it may age very slowly.  In particular, there are basins in trajectory space where initial conditions are remembered for all time.  Those corresponding to ordered crystals are often accessible at equilibrium conditions.  On the other hand, those associated with disordered solids are negligible at equilibrium but can become dominant in ensembles that are weighted away from equilibrium.  Once driven to a configuration of one of the members of that ensemble, atomic configurations of that system will be remembered for times far into the future.  These configurations are examples of non-ergodic glassy states.  They can be metastable with respect to crystal phases, but reaching a crystal from one of these disordered states requires nucleation processes --  creation of interfaces that grow to macroscopic sizes.  The creation of interfaces are necessarily much slower than structural reorganization, and in the amorphous solid state, structural relaxation already takes a very long time. 

There should be no controversy surrounding the existence of this broken symmetry in time, a phase transition between ergodic and non-ergodic phases.  It is a fact established with exact analysis of idealized models, and convincing numerical evidence for more realistic atomistic models.  The existence of this transition provides a natural explanation for the formation of glass and its equilibrium precursors, such as growing time scales, dynamic heterogeneity and intermittency, and it also explains why a glass and its melt are not easily distinguished by spatial structure alone.   Nevertheless, it remains unresolved whether this transition is most often or even ever the actual glass transition encountered in real non-equilibrium processing of glasses.  Resolution on this front must await the development of principles that link experimental protocols with those of re-weighting ensembles of trajectories, analogous to developments of the last century that linked experimental protocols to Gibbs' statistical mechanics.   We expect developments of this sort to begin to appear in the near future, possibly growing from the substantial and increasing body of molecular theory and experiment on driven systems, such as viscous fluids under shear \cite{Cates,Fielding,Olmsted}.

A theme of this review is the recommendation to think dynamically not thermodynamically, or if the latter, to include time into the system's dimensionality.  We find it remarkable that so many of the concepts of equilibrium phase transitions -- order parameters, broken symmetries, wetting, and so forth -- carry over to understanding order-disorder in space-time.   The concepts we have described here for the highly fluctuating equilibrium dynamics of glass formers can also be applied to explicitly non-equilibrium problems such as that of aging systems \cite{CuKu,Aging}.  Furthermore, this understanding of dynamics may likely extend to areas beyond glassy physics, such as nanoscale assembly \cite{Whitesides,Service}, protein folding dynamics and evolution \cite{Shak}, and signal transduction in biology \cite{Tyson}.  For example, signaling networks can be considered as many body problems of interacting proteins and protein complexes in the interior of the cell.  While the collection of proteins and complexes is disordered and ``crowded'' \cite{Crowding}, the interacting units in individual signalling pathways are sparse and the interactions are often catalytic, i.e. facilitated, which is bound to give rise to  fluctuation-dominated heterogeneous space-time structures analogous to those described above for the glass transition problem.  This is an issue that has only has started to be addressed \cite{Chakraborty,Dinner}.

\acknowledgments

We warmly thank our collaborators on the work described in this review: Ludovic Berthier,  Yael Elmatad, Lester Hedges, Rob Jack, YounJoon Jung,  David Kelsey,  Vivien Lecomte, Lutz Maibaum, Mauro Merolle, Albert Pan, Kristina van Duijvendijk, Fred van Wijland, and Steve Whitelam.  We are grateful to Lester Hedges and Lutz Maibaum for help in the preparation of the figures, and to Rob Jack, David Limmer and Thomas Speck for comments on the manuscript.  JPG was partly supported by EPSRC under grant No. GR/S54074/01. DC's research in this area  has been funded over the last decade by grants from the National Science Foundation, the Office of Basic Energy Sciences of the US Department of Energy, and the Office of Naval Research of the US Department of Defense.


\begin{thebibliography}{99}
\bibitem{Example} For example, consider liquid ortho-terphenyl (OTP), where $T_{\mathrm{g}}$ is in the range of 243K to 246K, and the onset temperature, $T_{\mathrm{o}}$, is about 350K~\cite{Elmatad}; extrapolation of thermodynamic data determines $T_{\mathrm{K}}$ to be in the range of 200K to 204K, while all but one of the various values of $T_{\mathrm{VFT}}$ extracted from fitting transport data differ substantially from $T_{\mathrm{K}}$:  $T_{\mathrm{VFT}}=184$K~\cite{Tanaka}, $T_{\mathrm{VFT}}=202$K~\cite{Angell-Sc}, $T_{\mathrm{VFT}}=211$K~\cite{Elmatad} and $T_{\mathrm{VFT}}=186$K~\cite{Elmatad}. Similarly, consider triphenyl phosphate (TPP), where $T_{\mathrm{g}}$ is 240K, and the onset temperature, $T_{\mathrm{o}}$, is 286K~\cite{Elmatad}; extrapolation of thermodynamic data determines $T_{\mathrm{K}}=166$K~\cite{Tanaka}, while fitting to transport data gives $T_{\mathrm{VFT}}=183$K~\cite{Tanaka} or $T_{\mathrm{VFT}}=177$K~\cite{Elmatad}.
\bibitem{Complexity}
Equation (\ref{Zq}) can be dominated by the inactive state when the number of inactive initial conditions is exponential in system size.  In that case the typical dynamical state becomes the inactive one.  This is what would appear to happen in certain systems with long range interactions \cite{Rob-Fred,Rob-ROM}.
\end{thebibliography}
\end{document}